\documentclass[12pt]{iopart}
\usepackage{epsf}

\begin{document}

\title{Fano resonances in a three-terminal nanodevice}
\author {V A Margulis and M A Pyataev}

\address{Institute of Physics and Chemistry,
Mordovian State University, 430000, Saransk, Russia}
\ead{theorphysics@mrsu.ru}

\begin{abstract}
The electron transport through a quantum sphere with three one-dimensional
wires attached to it is investigated.
An explicit form for the transmission coefficient
as a function of the electron energy is found
from the first principles.
The asymmetric Fano resonances are detected in transmission
of the system.
The collapse of the resonances is shown to appear
under certain conditions.
A two-terminal nanodevice with an additional gate lead
is studied using the developed approach.
Additional resonances and minima of transmission are
indicated in the device.
\end{abstract}


\section{Introduction}

Electron transport in nanoscale multiterminal ballistic devices
attracts considerable attention in the last decade.
Rapid advances in nanoelectronic fabrication techniques
have made possible the realization of
electron waveguide devices with dimensions smaller than the
elastic and inelastic scattering lengths of conduction electrons.
Various interesting multiterminal
nanoelectronic devices,
such as the single electron transistor \cite{GGG98,GGH00,ZGG01}
and the three-terminal ballistic junction or Y-branch switch
\cite{CX02, Xu02,CX03}
have been proposed
as a promising alternative for future low-power, high-speed
switching devices.
Recent theoretical studies reported
transistor-like behaviour of various three-terminal
molecule-based devices \cite{SB02}.

A number of theoretical and experimental works
has been focused on the investigation of the electron transport in
multiterminal quantum systems.
Three-terminal ballistic junctions were studied in \cite{Xu02,CX03}.
The electron transport in a three-terminal molecular
wire connected to metallic leads
was investigated in \cite{EK00}.

One of the interesting phenomena detected in these systems is
Fano resonances in the transmission probability.
Being a characteristic manifestation of wave phenomena
in a scattering experiment
resonances have received considerable attention in recent
electron transport investigations.
A number of papers
\cite{CWB01, BS01, XG01, THC02, ZCP02}
is devoted to the study of Fano resonances in the transport
through various quantum dots.
Resonant tunnelling  through quasi-one-dimensional
channels with impurities is
investigated in  \cite{KS99, KSJ99, KS99E, KSR02}.
The temperature dependence of the zero-bias conductance
of the single-electron transistor
is considered in \cite{ZGG01}.
Coherent transport through a quantum
dot embedded in an Aharonov-Bohm ring
is investigated in \cite{Kang99}.
Line shape of resonances
in the overlapping regime
is studied in \cite{MRS03}.

Interference phenomena closely related to the Fano resonances attract
considerable attention in the past few years.
Those resonances are of universal nature and have been observed in
various systems. We mention, for example, atom photoionization,
electron and ion scattering, Raman scattering and so on.
Recently, the line shape of resonances has been discussed
in experiments on electron transport through mesoscopic
systems \cite{GGH00, ZGG01, KAK02}.
It is shown in \cite{GMP03}  that the same resonances occur
in the electron transport through a quantum nanosphere
with two wires attached to it.

Recent progress in nanotechnology has made it possible to fabricate
conductive two-dimensional nanostructures with spherical symmetry
such as fullerenes and metallic spherical nanoshells.
A number of works is devoted to the theoretical study
of the electron transport on spherical surfaces \cite{FLP95, Kis97, BGMP02}.
The purpose of the present paper is an investigation
of the electron transport through
a three-terminal nanodevice consisting of
a conductive nanosphere ${\rm S}$ with
three one-dimensional wires attached to it
at the points ${\bi q}_j$ ($j=1\ldots3$).
We denote by ${\bi q}_j$
a set of spherical coordinates
$(\theta_j,\varphi_j)$ of the point.

\section{Hamiltonian and transmission coefficient}

\begin{figure}[!h]
\begin{center}
\epsfclipon
\epsfxsize=100mm
\epsfbox{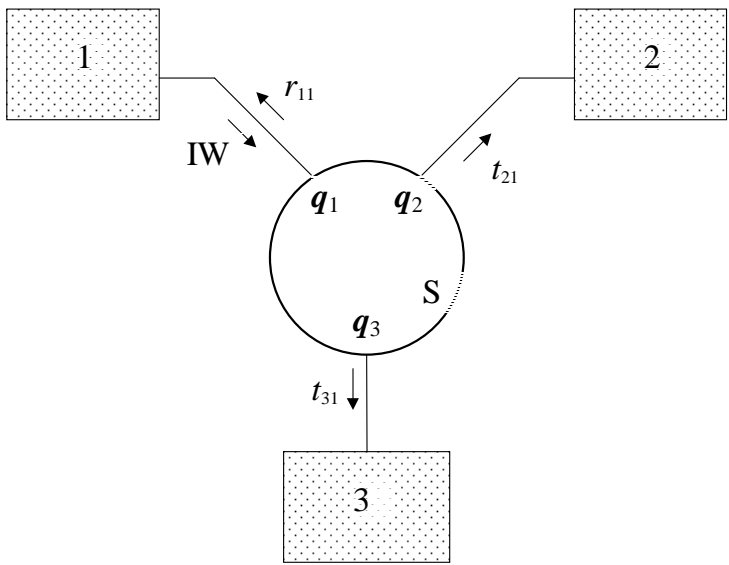}
\end{center}
\caption{\label{f-scheme}
Scheme of the device.
An incident wave (IW) originating from reservoir 1 is reflected back
with amplitude $r_{11}$ and scattered to reservoirs 2 and 3
with amplitudes $t_{21}$ and $t_{31}$ respectively.}
\end{figure}

In our model, the wires are taken to be one-dimensional
and represented by semiaxes ${\mathbf R}^+_j=\{x: x\geq 0\}$ ($j=1\ldots3$).
They are connected to the sphere by gluing the
point $x=0$ from ${\mathbf R}^+_j$ to the point ${\bi q}_j$ from ${\rm S}$.
We suppose ${\bi q}_i\neq{\bi q}_j $ for $i\neq j$.
The scheme of the device is shown in
figure~\ref{f-scheme}.
Here $t_{21}(E)$ and $t_{31}(E)$ are the transmission
amplitudes of the electron wave
and $r_{11}(E)$ is the reflection amplitude.

The Hamiltonian of
a free electron in the wire is $H_j=p_x^2/2{m^*}$,
where ${m^*}$ is the electron effective mass and $p_x$ is the
momentum operator for the electron in wires.
Electron motion on the sphere is described by the Hamiltonian
$H_{\rm S}={\bi L}^2/2{m^*}{r}^2$ where
$r$ is the radius of the nanosphere
and ${\bi L}$ is the angular momentum operator.
A wavefunction $\psi$ of the electron in the device consists
of four parts: $\psi_{\rm S}$, $\psi_1$, $\psi_2$ and $\psi_3$,
where $\psi_{\rm S}$ is a function on ${\rm S}$ and
$\psi_j$ ($j= 1\ldots 3$) are functions on ${\mathbf R}^+_j$.
We note that in general case $\psi_{\rm S}$ is not the eigenfunction of
the operator $H_{\rm S}$.

The Hamiltonian $H$ of the whole system is
a point perturbation of the operator
\begin{equation}
H_0=H_{\rm S}\oplus H_1\oplus H_2\oplus H_3.
\end{equation}
To define this perturbation we use boundary conditions at points of gluing.
The role of boundary values for the wavefunction $\psi_j(x)$
is played, as usual, by $\psi_j(0)$ and $\psi_j'(0)$.
The zero-range potential theory shows
that to obtain a non-trivial Hamiltonian
on the whole system we must consider
functions $\psi_{\rm S}({\bi x})$ with a logarithmic
singularity at points of gluing ${\bi q}_j$ \cite{BG03}
\begin{equation}
            \label{asymp}
\psi_{\rm S}({\bi x})=
-u_j{{m^*} \over \pi \hbar^2}\ln\rho({\bi x},{\bi q}_j)+v_j +o(1)
\end{equation}
as ${\bi x} \to {\bi q}_j$.
Here $u_j$ and $v_j$ are complex coefficients and
$\rho({\bi x},{\bi q})$ is the geodesic distance
on the sphere between the points
${\bi x}$ and ${\bi q}_j$.
It is known that the most general
self-adjoint boundary conditions are defined by some
linear relations between
$\psi_j(0)$, $\psi'_j(0)$ and the coefficients
$u_j$, $v_j$.
Following \cite{GMP03} we will write this conditions
in the form
\begin{equation}
                       \label{bound}
\cases{v_j=\sum\limits_{k=1}^{3}\left[ B_{jk}u_k-
(\hbar^2/2{m^*})
A_{jk }\psi'_k(0)\right], \cr
\psi_j(0)=\sum\limits_{k=1}^{3}\left[ A_{kj}^* u_k-
(\hbar^2/2{m^*})C_{jk}\psi'_k(0)\right], & ${j=1\ldots 3.}$ \cr}
\end{equation}
Here complex parameters $A_{jk}$, $ B_{jk}$ and $C_{jk}$  form
$3\times3$ matrices.
The matrices $B$ and $C$ have to be Hermitian because
the Hamiltonian $H$ is a self-adjoint operator \cite{BG03}.
To avoid a non-local tunnelling coupling  \cite{BGMP02}
between different contact points
we will restrict ourselves to the case of diagonal matrices
$A_{jk}$, $ B_{jk}$ and $C_{jk}$ only.

According to the zero-range potential theory diagonal elements
of the matrix $B$ determine the strength of point perturbations
of the Hamiltonian $H_{\rm S}$ at the
points ${\bi q}_j$ on ${\rm S}$ \cite{BGMP02}.
These elements may be expressed in terms of scattering lengths
$\lambda^B_j$ on the
corresponding point perturbations:
$B_{jj}={m^*}\ln(\lambda^B_j)/\pi\hbar^2$.
Similarly, elements $C_{jj}$  describe the strength of point perturbations
at the points $x=0$ in the wires and may be
expressed in terms of scattering lengths $\lambda^C_j$
by the relation
$C_{jj}= -{m^*}\lambda^C_j/2\hbar^2$ \cite{GMP03}.
For convenience, we represent parameters $A_{jj}$ in the form
$A_{jj}={m^*}\sqrt{\lambda^A_j}\rme^{\rmi\phi_j}/\hbar^2$,
where $\lambda^A_j$ has the dimension of length
and $\phi_j$ is the argument of the complex number $A_{jj}$.
Note that the effect of the scattering lengths
$\lambda^A_j$, $\lambda^B_j$ and $\lambda^C_j$
on the electron transport has been discussed in \cite{GMP03}.
In the present paper we concentrate our attention on the facts
independent of the contact parameters.

To obtain transmission and reflection
coefficients of the system one needs a solution of
the Schr\"odinger equation for the Hamiltonian $H$.
The function $\psi_{1}(x)$ in this solution is a superposition
of incident and reflected waves
while the functions $\psi_{2}(x)$ and $\psi_{3}(x)$
represent scattered waves.
The wavefunction $\psi_{\rm S}({\bi x})$ may be expressed in terms of
the Green function $G({\bi x},{\bi y};E)$ of the Hamiltonian $H_{\rm S}$
\cite{GMP03}
\begin{equation}
        \label{solution}
\cases
{\psi_{\rm S}({\bi x})=\sum\limits_{j=1}^3\xi_j(E) G({\bi x},{\bi q}_j;E), \cr
\noalign{\medskip}
\psi_1(x)= \rme^{-\rmi kx}+r_{11}(E)\rme^{\rmi kx},\cr
\noalign{\medskip}
\psi_2(x)= t_{21}(E)\rme^{\rmi kx},\cr
\noalign{\medskip}
\psi_3(x)= t_{31}(E)\rme^{\rmi kx}.\cr}
\end{equation}
Here $k=\sqrt{2{m^*} E/\hbar^2}$ is the electron wave vector in wires and
$\xi_j(E)$ are complex factors.

It is well known \cite{GS} that the Green function
$G({\bi x},{\bi y};E)$ may be expressed in the form
\begin{equation}
G({\bi x},{\bi y};E)=-{{m^*}\over2\hbar^2}{1\over \cos(\pi t)}
{\cal P}_{t-{\case12}}\left(
-\cos\left({\rho({\bi x},{\bi y})/{r}}\right)\right)
\end{equation}
where ${\cal P}_{\nu}(x)$ is the Legendre function
and $t(k)=\sqrt{{r}^2k^2+1/4}$.

Considering the asymptotics (\ref{asymp}) of $\psi_{\rm S}({\bi x})$
from (\ref{solution})
near the point ${\bi q}_j$, we have
\[
u_j=\xi_j(E),\qquad
v_j=\sum\limits_{i=1}^{3}Q_{ij}(E)\xi_i(E)\,.
\]
Here $Q_{ij}(E)$ is the so-called Krein's ${\cal Q}$-function,
that is $3\times 3$ matrix with elements
\begin{equation}
Q_{ij}(E)=\cases{
G({\bi q}_i ,{\bi q}_j ;E), &  $i\ne j$;\cr
\noalign{\medskip}
\displaystyle\lim\limits_{{\bi x} \to {\bi q}_j}
\left[
G({\bi q}_j ,{\bi x} ;E)
+\frac{{m^*}}{\pi\hbar^2} \ln \rho({\bi q}_j,{\bi x})
\right]\,, &  $i=j$.\cr}
\end{equation}
Using the asymptotic expression for the Legendre function in a vicinity
of the point $x=-1$, we get the following form for diagonal elements of
${\cal Q}$-matrix \cite{BGMP02}
\begin{equation}
\fl
Q_{jj}(E)=-{{m^*}\over\pi\hbar^2}
\left[\Psi\left(t(k)+\case12\right)-
{\pi\over2}\tan(\pi t(k))-\ln(2{r})+C_{\rm E} \right],\qquad j=1\ldots 3
\end{equation}
where $\Psi(x)$ is the logarithmic derivative of the $\Gamma$-function
and $C_{\rm E}$ is the Euler constant.

Substituting (\ref{solution}) into (\ref{bound}),
we get a system of six linear equations for
$\xi_j$, $r_{11}$, $t_{21}$ and $t_{31}$.
For convenience, we introduce dimensionless
elements of ${\cal Q}$-matrix
\[
{\widetilde Q}_{ij}(E)=(\hbar^2/{m^*})({\cal Q}_{ij}(E)-B_{ij}).
\]
Solving the system of equations, we obtain
\begin{equation}
r_{11}=\frac{(k\lambda^C_1-4\rmi)\Delta_1}{(k\lambda^C_1+4\rmi)\Delta}
\end{equation}
where
\begin{equation}
\Delta=\left|
    \begin{array}{c c c}
    \displaystyle {\widetilde Q}_{11}-{2k\lambda^A_{1}\over k\lambda^C_1+4\rmi }&{\widetilde Q}_{12}&{\widetilde Q}_{13}\\
    {\widetilde Q}_{21}&{\widetilde Q}_{22}-
    \displaystyle{2k\lambda^A_{2}\over k\lambda^C_2 + 4\rmi }&{\widetilde Q}_{23}\\
    {\widetilde Q}_{31}&{\widetilde Q}_{32}&{\widetilde Q}_{33}-
    \displaystyle{2k\lambda^A_{3}\over k\lambda^C_3+ 4\rmi}\\
    \end{array}
\right|,
\end{equation}
and
\begin{equation}
\Delta_1=\left|
    \begin{array}{c c c}
    \displaystyle {\widetilde Q}_{11}-{2k\lambda^A_{1}\over k\lambda^C_1-4\rmi }&{\widetilde Q}_{12}&{\widetilde Q}_{13}\\
    {\widetilde Q}_{21}&{\widetilde Q}_{22}-
    \displaystyle{2k\lambda^A_{2}\over k\lambda^C_2 + 4\rmi }&{\widetilde Q}_{23}\\
    {\widetilde Q}_{31}&{\widetilde Q}_{32}&{\widetilde Q}_{33}-
    \displaystyle{2k\lambda^A_{3}\over k\lambda^C_3+ 4\rmi}\\
    \end{array}
\right|.
\end{equation}
The transmission amplitude $t_{21}$ is given by
\begin{equation}
        \label{trans}
\fl
t_{21}=\frac{16k\sqrt{\lambda^A_1\lambda^A_2}
\rme^{\rmi(\phi_1-\phi_2)}
\left[{2k\lambda^A_3}{\widetilde Q}_{21}-(k\lambda^C_3+4\rmi)
({\widetilde Q}_{21}{\widetilde Q}_{33}-
{\widetilde Q}_{23}{\widetilde Q}_{31})
\right]}
{\rmi(k\lambda^C_1+4\rmi)(k\lambda^C_2+4\rmi)(k\lambda^C_3+4\rmi)\Delta}.
\end{equation}
Similarly, we can write
\begin{equation}
\fl
t_{31}=\frac{16k\sqrt{\lambda^A_1\lambda^A_3}
\rme^{\rmi(\phi_1-\phi_3)}
\left[{2k\lambda^A_2}{\widetilde Q}_{31}-(k\lambda^C_2+4\rmi)
({\widetilde Q}_{31}{\widetilde Q}_{22}-
{\widetilde Q}_{32}{\widetilde Q}_{21})
\right]}
{\rmi(k\lambda^C_1+4\rmi)(k\lambda^C_2+4\rmi)(k\lambda^C_3+4\rmi)\Delta}.
\end{equation}
We emphasize that the relation
\begin{equation}
|r_{11}|^2+|t_{21}|^2+|t_{31}|^2=1
\end{equation}
is valid for arbitrary energy $E$ in compliance with the current
conservation law.

\begin{figure}[!h]
\begin{center}
\epsfclipon
\epsfxsize=130mm
\epsfbox{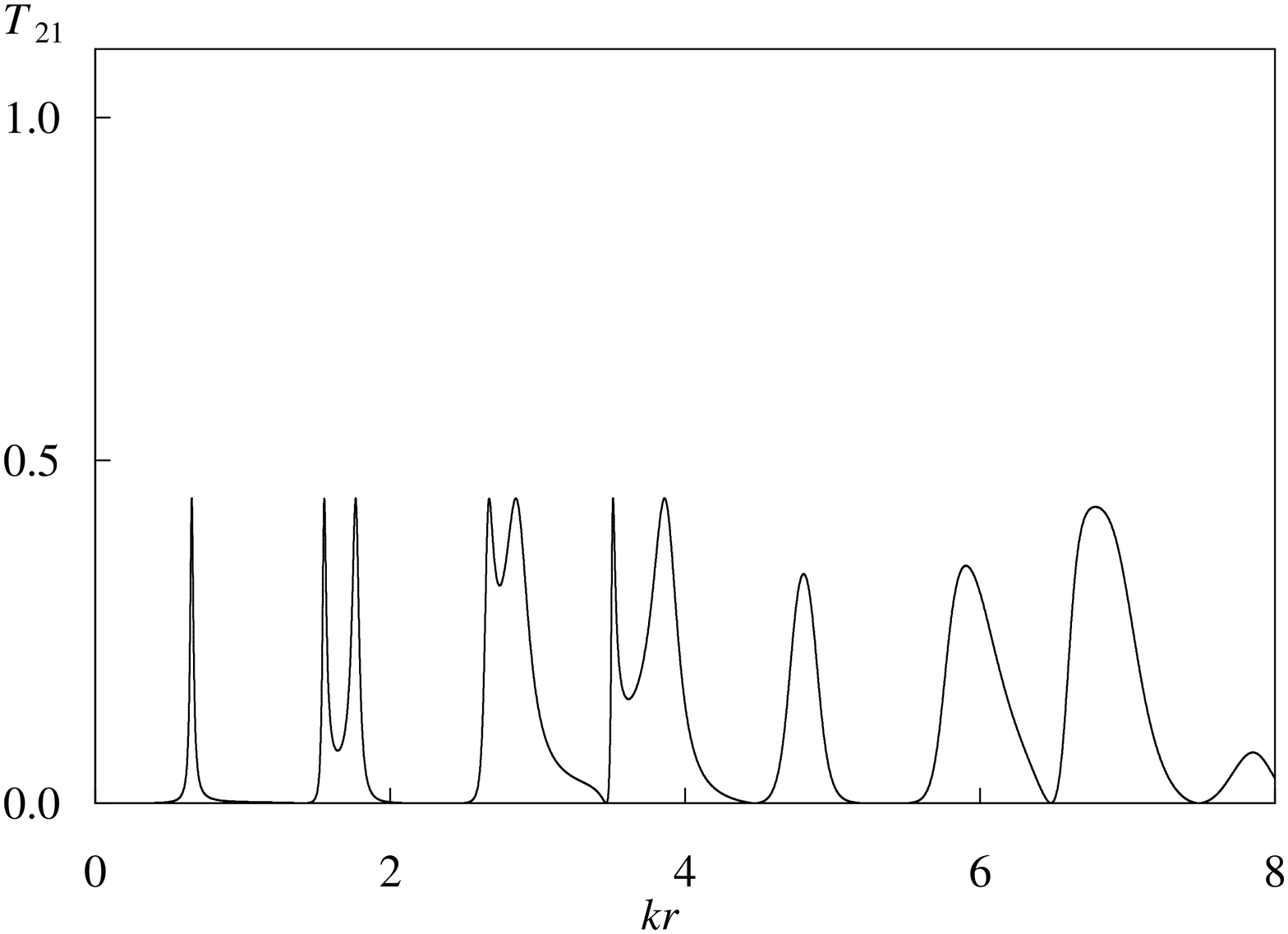}
\end{center}
\caption{\label{f-symm}
Transmission coefficient
$T_{21}$ as a function of the dimensionless parameter $k{r}$
at $\rho_{12}=\rho_{13}=\rho_{23}=\case23\pi{r}$.}
\end{figure}

The transmissions coefficient $T_{21}\equiv|t_{21}|^2$
as a function of the dimensionless
parameter $k{r}$ is shown in figure~\ref{f-symm}.
The figure corresponds to the case when
contacts are placed equidistant on the great circle of the sphere.
Denoting by $\rho_{ij}$ the distance $\rho({\bi q}_i,{\bi q}_j)$
between the points ${\bi q}_i$ and ${\bi q}_j$, we can represent
the position of contacts by relation
$\rho_{12}=\rho_{13}=\rho_{23}=2\pi{r}/3$.
In this case, the relation $t_{21}=t_{31}$ is valid for arbitrary energy
due to the symmetry of system.
Therefore
$T_{21}$ does not exceed the value $\case12$.
All figures correspond to the case
$\lambda^A_j=\lambda^B_j=\lambda^C_j=0.1 {r}$ for all $j$.

\section{Fano resonances}

It is evident from equation (\ref{trans}) that the transmission
amplitude $t_{21}(E)$ has zeros of two different types.
The zeros of the first type
are stipulated by  the poles of $Q_{ij}(E)$ and
coincide with the eigenvalues $E_l$
of the operator $H_{\rm S}$.
The denominator in (\ref{trans}) has a pole of
the third order at $E=E_l$ while the numerator has a pole of
the second order only. Hence, the transmission coefficient vanishes
in these points.

The zeros of the second type are determined by the following equation:
\begin{equation}
2k\lambda^A_3 {\widetilde Q}_{21} - (k\lambda^C_3+4\rmi)
({\widetilde Q}_{21}{\widetilde Q}_{33}-{\widetilde Q}_{23}{\widetilde Q}_{31})=0.
\end{equation}
The positions of the second-type zeros depend on the arrangement
of ${\bi q}_j$ on the sphere
in contrast to the positions of the first-type zeros.

We will show below that
in a vicinity of the first-type zeros $E_l$
transmission coefficient has a form of
the asymmetric Fano resonance.
Consider the form of $Q_{ij}(E)$ near the point $E_l$
\begin{equation}
            \label{asympQ}
Q_{ij}(E)\simeq \frac{\alpha_{ij}}{E_l-E} +\beta_{ij}.
\end{equation}
The residues $\alpha_{ij}$ of $Q_{ij}(E)$ at the point
$E_l$ may be expressed in terms of eigenfunctions
of the operator $H_{\rm S}$
\begin{equation}
\alpha_{ij}=\sum\limits_{m=-l}^{l}
Y_{lm}({\bi q}_i){Y_{lm}^*({\bi q}_j)}
\end{equation}
where $Y_{lm}({\bi x})$ are the spherical harmonics.

Denote by $\tilde\beta_{ij}$ modified matrix $\beta$
\[
\tilde\beta_{ij}=\beta_{ij}-B_{ij}-
\frac{2{m^*} k\lambda^A_j}{\hbar^2 (k\lambda^C_j +4\rmi)}\delta_{ij}\,.
\]
Substituting (\ref{asympQ}) into (\ref{trans})
and considering linear in $E-E_l$ approximation for
the numerator and the denominator of (\ref{trans}), we obtain
\begin{equation}
            \label{Fano}
t_{21}(E)\simeq \eta\frac{E-E_l}{E-E_R-\rmi\Gamma}\,.
\end{equation}
Here $E_R$ determines the position of the asymmetric peak,
$\Gamma$ is the half-width of the resonance,
and $\eta$ is a normalization factor.
It is evident from (\ref{Fano}) that the transmission coefficient
has a form of the Fano resonance near $E_l$.
The parameters $E_R$ and $\Gamma$ of the Fano resonance
are determined by
\begin{equation}
            \label{ER}
E_R+\rmi\Gamma=E_l+
\frac {\det \alpha}
{\sum\limits_{i,j}[\alpha_{ij}]^c\tilde\beta_{ij}}
\end{equation}
where $[\alpha_{ij}]^c$ is the algebraic complement of $\alpha_{ij}$
in the matrix $\alpha$.
The normalization factor $\eta$ is given by
\begin{equation}
            \label{eta}
\eta=\frac{16{m^*}k\sqrt{\lambda^A_1\lambda^A_2}
\exp(\rmi(\phi_1-\phi_2))}
{\rmi \hbar^2(k\lambda^C_1+4\rmi)(k\lambda^C_2+4\rmi)
\sum\limits_{i,j}[\alpha_{ij}]^c \tilde\beta_{ij}}
(\alpha_{23}\alpha_{31}-\alpha_{21}\alpha_{33}).
\end{equation}
Note that the asyptotics (\ref{Fano}) for $t_{21}(E)$
is valid for arbitrary parameters of contacts $\lambda^A_j$,
$\lambda^B_j$ and $\lambda^C_j$.

\begin{figure}[!h]
\begin{center}
\epsfclipon
\epsfxsize=130mm
\epsfbox{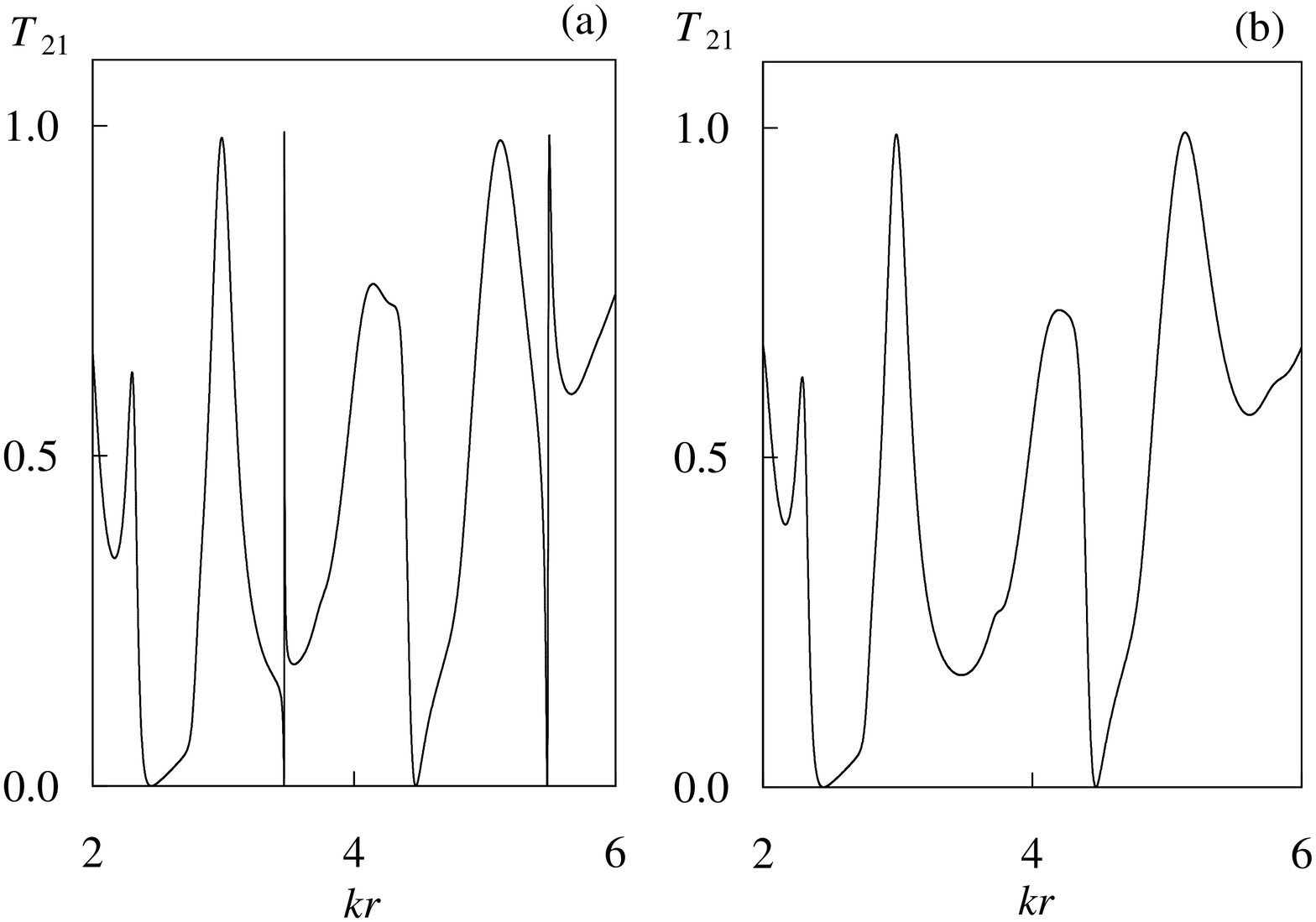}
\end{center}
\caption{\label{f-collapse}
Transmission coefficient $T_{21}$
as a function of the dimensionless parameter $k{r}$.
(a) $\rho_{12}=0.98\pi r$, $\rho_{13}=0.52\pi r$,
$\rho_{23}=0.5\pi r$;
(b) $\rho_{12}=\pi r$, $\rho_{13}=\rho_{23}=0.5\pi r$ (collapse of the Fano resonances).}
\end{figure}

If $\det\alpha=0$ at a given $l$ then a collapse of
the Fano resonance occurs near $E_l$.
In this case, the pole and the zero of the
transmission amplitude coincide and cancel each other
(figure~\ref{f-collapse}).
Note that the second-type zeros remain on the plot of $T_{21}(E)$
in contrast to the situation considered in \cite{GMP03,BGMP02}.

To define the condition of the collapse we introduce
three complex vectors ${\bi V}_j$ by the following equation
\[
{\bi V}_j=\left(
\begin{array}{c}
   Y_{l,l}({\bi q}_j) \\
   Y_{l,l-1}({\bi q}_j)\\
   {\ldots} \\
   Y_{l,-l}({\bi q}_j) \\
\end{array}
\right).
\]
Matrix $\alpha$ is the Gram matrix for vectors ${\bi V}_j$
because $\alpha_{ij}=\langle {\bi V}_i|{\bi V}_j\rangle$.
Hence, the condition $\det\alpha =0$ is satisfied if and only if
vectors ${\bi V}_j$ are linearly dependent.

If we choose
a coordinate system on the sphere so that the points
${\bi q}_j$ were on the circle $\theta=\theta_0=const$
and fix the origin of the azimuthal angle $\varphi$
at the point ${\bi q}_1$,
then points ${\bi q}_j$ have the following coordinates
\[
{\bi q}_1=(\theta_0, 0),\qquad
{\bi q}_2=(\theta_0, \varphi_2),\qquad
{\bi q}_3=(\theta_0, \varphi_3).
\]
Vectors ${\bi V}_j$ can be represented in the form
\[
{\bi V}_1=\left(
\begin{array}{l}
   f_l^{l} \\
   f_l^{l-1} \\
   {\ldots}\\
   f_l^{-l}\\
\end{array}
\right),\qquad
{\bi V}_2=\left(
\begin{array}{l}
   f_l^{l}\rme^{\rmi l\varphi_2} \\
   f_l^{l-1}\rme^{\rmi (l-1)\varphi_2} \\
   {\ldots} \\
   f_l^{-l}\rme^{-\rmi l\varphi_2} \\
\end{array}
\right),\qquad
{\bi V}_3=\left(
\begin{array}{l}
   f_l^{l}\rme^{\rmi l\varphi_3} \\
   f_l^{l-1}\rme^{\rmi (l-1)\varphi_3} \\
   {\ldots} \\
   f_l^{-l}\rme^{-\rmi l\varphi_3} \\
\end{array}
\right)
\]
where $f_l^m=C_{ml}P_l^{|m|}(\cos\theta)$,
$P_l^{|m|}(x)$ are the Legendre polynomials, and $C_{ml}$ are
the normalization factors of the spherical harmonics.

Denote by $M$ the $3\times(2l+1)$ matrix
composed of three vectors ${\bi V}_j$.
The condition $\det\alpha\neq0$ holds if and only if
rank of the matrix $M$ is $3$.
In general, if points ${\bi q}_j$
are placed on the sphere in random manner,
all vectors ${\bi V}_j$  are linearly independent.
If $\varphi_2=\pi$ then all elements of $M$ with different parity
of $m$ and $l$ are equal zero
since $\theta_0=\pi/2$ and
$P_l^{|m|}(0)=0$ for odd $m+l$.
Elements of ${\bi V}_{2}$ with even $l+m$
in this case are equal $(-1)^l f_l^{m}$.
Hence the condition ${\bi V}_2=(-1)^l{\bi V}_1$
is satisfied that directly implies $\det \alpha=0$.
Thus the collapse of the Fano resonances takes place
if the points ${\bi q}_1$ and ${\bi q}_2$ are antipodal
on the sphere.
This condition is independent of the position of ${\bi q}_3$.

It is evident that condition $\det \alpha=0$ holds
if any pair of three points ${\bi q}_j$
consists of antipodal points.
But if $\varphi_2=\pi$ or $|\varphi_2-\varphi_3|=\pi$
then the normalization factor $\eta$
vanishes because $\alpha_{23}\alpha_{31}-\alpha_{21}\alpha_{33}=0$.
In this case, the linear approximation
for the denominator and the numerator
of $t_{21}$ is inapplicable, and
quadratic in $E-E_l$ terms in (\ref{trans}) must be taken into account.
The equation similar to (\ref{Fano}) may be obtained
for $t_{21}$ with
\begin{equation}
E_R+\rmi\Gamma=E_l+
{\sum\limits_{i,j}[\alpha_{ij}]^c\tilde\beta_{ij}}
\left(\sum\limits_{i,j}\alpha_{ij}[\tilde\beta_{ij}]^c\right)^{-1}.
\end{equation}
In this case, the half-width $\Gamma$ of the resonance
is determined by the parameters $\tilde \beta_{ij}$, and
the condition $\Gamma=0$ requires a special choice of
scattering lengths $\lambda^A_j$, $\lambda^B_j$ and $\lambda^C_j$.
Therefore, in general, the collapse of the Fano resonances
appears when the points ${\bi q}_1$ and ${\bi q}_2$ only
are antipodal on the sphere.

\section{Two-terminal device with an additional gate lead}

\begin{figure}[!h]
\begin{center}
\epsfclipon
\epsfxsize=100mm
\epsfbox{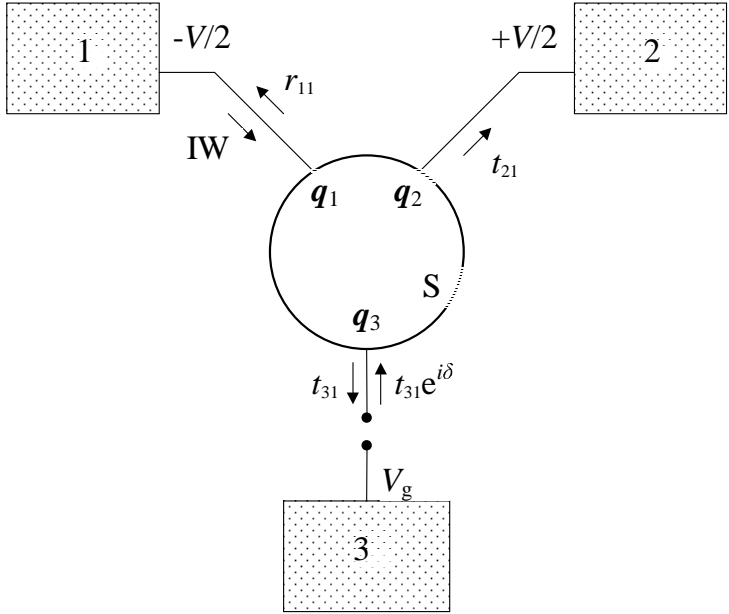}
\end{center}
\caption{\label{f-scheme2}
Scheme of the nanodevice with the break in the third wire.
$V$ is the bias voltage between reservoirs 1 and 2 and $V_g$ is the gate voltage.}
\end{figure}

The dependence $T_{21}(E)$ is of particular interest because
according to the Landauer--B\"uttiker formula
the conductance of the system
as a function of the chemical potential has the same form at zero
temperature.
For experimental observation of such a dependence one
needs to change the electrochemical potential
of electrons  on the
sphere relative to the Fermi energies in reservoirs.
This may be realized by using an additional gate electrode near the
sphere which is connected to the system through a potential barrier.
Here we consider this additional gate lead
as one-dimensional broken wire.
The scheme of the studied device is shown in figure~\ref{f-scheme2}.
In this case, one can shift energy levels of electrons on the sphere
relative to the Fermi energy in the reservoirs 1 and 2
by changing the voltage $V_{g}$.

The electron wave outgoing from the sphere in this case
reflects in the third wire and returns back completely.
The solution of the Schr\"odinger equation for this system
differs from (\ref{solution}) by the expression for $\psi_3(x)$
\begin{equation}
\psi_3(x)= t_{31}\rme^{\rmi kx}+t_{31}\rme^{\rmi\delta}\rme^{-\rmi kx}
\end{equation}
where $\delta=2kL+\pi$ is the phase incursion and
$L$ is the distance between ${\bi q}_3$ and the point of break.

The transmission coefficient in this case may be expressed in the form
\begin{equation}
\fl
t_{21}=\frac{16k\sqrt{\lambda^A_1\lambda^A_2}
\rme^{\rmi(\phi_1-\phi_2)}
\left[{2k\lambda^A_3}{\widetilde Q}_{21}-(k\lambda^C_3-4\cot(\delta/2))({\widetilde Q}_{21}{\widetilde Q}_{33}-{\widetilde Q}_{23}{\widetilde Q}_{31})
\right]}
{\rmi(k\lambda^C_1+4\rmi)(k\lambda^C_2+4\rmi)(k\lambda^C_3-4\cot(\delta/2))\tilde\Delta}
\end{equation}
where
\begin{equation}
\tilde\Delta=\left|
    \begin{array}{c c c}
    \displaystyle {\widetilde Q}_{11}-{2k\lambda^A_{1}\over k\lambda^C_1+4\rmi }&{\widetilde Q}_{12}&{\widetilde Q}_{13}\\
    {\widetilde Q}_{21}&{\widetilde Q}_{22}-
    \displaystyle{2k\lambda^A_{2}\over k\lambda^C_2 + 4\rmi }&{\widetilde Q}_{23}\\
    {\widetilde Q}_{31}&{\widetilde Q}_{32}&{\widetilde Q}_{33}-
    \displaystyle{2k\lambda^A_{3}\over k\lambda^C_3-4\cot(\delta/2)}\\
    \end{array}
\right|.
\end{equation}

\begin{figure}[!h]
\begin{center}
\epsfclipon
\epsfxsize=130mm
\epsfbox{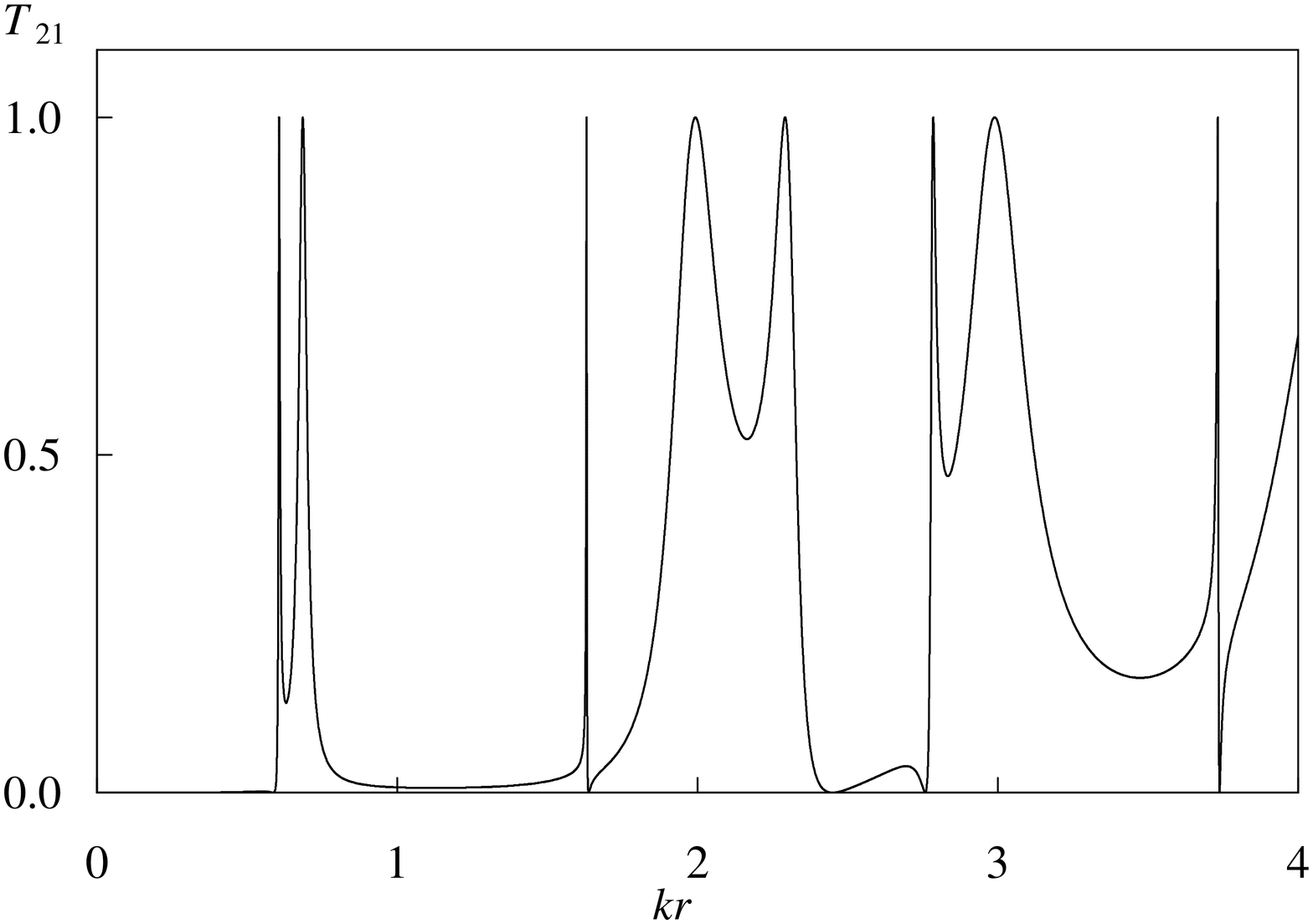}
\end{center}
\caption{\label{f-break}
Transmission coefficient as a function of the dimensionless parameter
$k{r}$ in case of broken third wire at
$\rho_{12}=\pi r$, $\rho_{13}=\rho_{23}=0.5\pi r$ and $L=0.57 r$.}
\end{figure}

The dependence $T_{21}(E)$ for the case of broken third wire
is shown in figure~\ref{f-break}.
In contrast to the case considered above
the height of peaks can reach a unity
since there is no energy loss due to the
outgoing of electrons into the third wire.
Moreover the additional resonance peaks and minima arise
because of the interference of electron waves in the broken wire.

\section{Conclusions}

Electron transport through a three-terminal nanodevice
is considered. Transmission and reflection coefficients of the device is found
by solving the Schr\"odinger equation.
We have shown that,
in general case, the function $T_{21}(E)$ has zeros of two different types
discussed in section 3.
Zeros of the first type coincide with the eigenvalues $E_l$ of unperturbed
electron Hamiltonian $H_{\rm S}$ on the sphere.
The transmission coefficient $T_{21}(E)$
has a form of asymmetric Fano resonance in a vicinity of the first-type zeros.
The parameters of the resonance
$E_R$ and $\Gamma$ are determined by equation (\ref{ER}).
If the points of contact ${\bi q}_1$ and ${\bi q}_2$
are placed antipodal on the sphere then the collapse of the Fano resonances
occurs.
In this case, the first-type zeros disappear while the second-type zeros remain
on the plot of $T_{21}(E)$ in contrast to the situation discussed in \cite{GMP03}.

Using the developed approach we consider the
two-terminal nanodevice with the additional gate electrode.
Additional resonances and minima of transmission
arise because of the interference of
electron waves in the third wire.

\ack
This work is financially supported by Russian Ministry of Education
(Grant A03-2.9-7).

\end{document}